# Molecular-scale Hydrophilicity Induced by Solute: Molecular-thick Charged Pancakes of Aqueous Salt Solution on Hydrophobic Carbon-based Surfaces

Guosheng Shi[1†], Yue Shen[1,2†], Jian Liu[1,3], Chunlei Wang[1], Ying Wang[1], Bo Song[1], Jun Hu[1], and Haiping Fang[1*]

[1]*Division of Interfacial Water and Key Laboratory of Interfacial Physics and Technology, Shanghai Institute of Applied Physics, Chinese Academy of Sciences, Shanghai 201800, China*

[2]*Salt Lake Resources and Chemistry Laboratory, Qinghai Institute of Salt Lakes, Chinese Academy of Sciences, Xining 810008, China*

[3]*University of Chinese Academy of Sciences, Beijing 100049, China*

†These authors contributed equally to this work.

*Email: fanghaiping@sinap.ac.cn

**ABSTRACT:** We directly observed molecular-thick aqueous salt-solution pancakes on a hydrophobic graphite surface under ambient conditions employing atomic force microscopy. This observation indicates the unexpected molecular-scale hydrophilicity of the salt solution on graphite surfaces, which is different from the macroscopic wetting property of a droplet standing on the graphite surface. Interestingly, the pancakes spontaneously displayed strong positively charged behavior. Theoretical studies showed that the formation of such positively charged pancakes is attributed to cation–π interactions between $Na^+$ ions in the aqueous solution and aromatic rings on the graphite surface, promoting the adsorption of water molecules together with cations onto the graphite surface; i.e., $Na^+$ ions as a medium adsorbed to the graphite surface through cation–π interactions on one side while at the same time bonding to water molecules through hydration interaction on the other side at a molecular scale. These findings suggest that actual interactions regarding carbon-based graphitic surfaces including those of graphene, carbon nanotubes, and biochar may be significantly different from existing theory and they provide new insight into the control of surface wettability, interactions and related physical, chemical and biological processes.

## INTRODUCTION

Hydrophilic/hydrophobic interactions are among the most important driving forces of various physical phenomena,[1–25] such as wetting/dewetting,[1–5] the folding and native structure formation of protein,[6–9] drug/molecular delivery,[10,11] water purification/desalination,[12,13] molecular recognition[14] and nanoparticle assembly/self-assembly in aqueous solution.[15,16] Conventionally the wetting property of a surface is determined by the macroscopic behavior of water, i.e., the contact angle of water droplets on the surface, and is widely used to analyze the interactions of the solid surface with other materials and dynamic properties at the interface. However, these interactions and dynamic properties are in fact dominated by the behavior of films of water molecules of molecular thickness on the surface. In recent years, molecular-thickness aqueous films have been observed on various hydrophilic surfaces[1–4] and between two surfaces.[17] In contrast, on typical hydrophobic carbon-based surfaces, such as graphene/graphite surfaces, water films are known to only adsorb at extremely low temperature.[2]

Carbon-based surfaces widely exist as surfaces of both nanoscale and macroscopic materials.[26–34] Most of these carbon-based surfaces, such as graphite, graphene,[27,28] carbon nanotubes,[29,30] fullerence,[31] biochar[32,33] and activated carbon,[34] contain a graphitic surface composed of many aromatic rings, which are hexagonal carbon rings rich in $\pi$ electrons. Moreover, graphitic surfaces rich in aromatic rings widely exist in biomolecules and organic molecules,[35] humus in the soil,[36] and polycyclic aromatic hydrocarbons in air pollutants.[37] It is well recognized that most of these graphitic surfaces rich in aromatic rings are hydrophobic and have wetting properties similar to those of graphite. However, some phenomena that only occur on hydrophilic surfaces have been recently observed on these graphitic carbon-based surfaces.[32–34,38–40] For example, biochar significantly increases water retention in a sandy soil even though it is well recognized as being hydrophobic.[32,33] The wetting and charge characteristics of graphene and graphitic surfaces are significantly impacted by the deposited substrates[38,39] and adsorbed airborne/atmospheric contaminants.[40,41] As stated above, the water wetting properties of graphitic surfaces are complicated and still far from being fully understood.

Characteristics of the interface between the graphitic surface and its surroundings, especially water and ionic environments, are believed to be central to the properties and application of carbon-based nanomaterials with graphitic surfaces.[27–45] We note that although the aromatic rings only weakly interact with water, they strongly interact with cations, in what is referred to as cation–$\pi$ interaction.[46] We thus expect that the behavior of the water close to the aromatic rings is strongly affected by the presence of cations.[47,48] However, in the study of the wetting property of surfaces, the role of cation–$\pi$ interactions has been ignored in spite of its importance in the control of the structure and function of microscale and nanoscale materials, macromolecules and proteins as has been extensively exploited.[35,49–51]

In this paper, we directly observe molecular-scale aqueous salt-solution pancakes on typical hydrophobic carbon-based surfaces, namely graphite surfaces, under ambient conditions at room temperature. We note that molecular-thick liquid films have been observed experimentally on a hydrophilic mica surface[1,2] and that water films are known to only adsorb on graphitic surfaces (including graphene) below ~150 K.[2] Our observation indicates unexpected molecular-scale hydrophilicity to the salt solution on the hydrophobic surfaces since macroscopic large salt-solution droplets can still form on the same surfaces. The cation–π interactions and the different properties of the graphitic surface and $Na^+$ ion distribution between molecular and macroscopic scales result in an inconsistency in the wetting property between molecular (hydrophilicity) and macroscopic (hydrophobicity) scales. Moreover, the pancakes spontaneously reveal strong positively charged behavior. The findings suggest extensive applications relating to biomolecules, treatment of heavy metal ion pollution in soil or water, and ion storage and detection.

**RESULTS AND DISCUSSION**

We deposited a drop of NaCl solution or pure water of millimeter dimensions onto a freshly cleaved highly oriented pyrolytic graphite (HOPG) sheet. In both cases, large macroscopic droplets formed on the graphite surface. The static contact angles (SCAs) of the droplets of NaCl solution and pure water on the HOPG surface were respectively measured as 95°± 4°(see Fig. 1D) and 93°± 2°(see Fig. S1D). Aqueous droplets were then blown from the surface with air (see schematic drawings Fig. 1(A–C) for NaCl solution and Supporting Information for pure water). The systems were at room temperature and ~40% relative humidity.

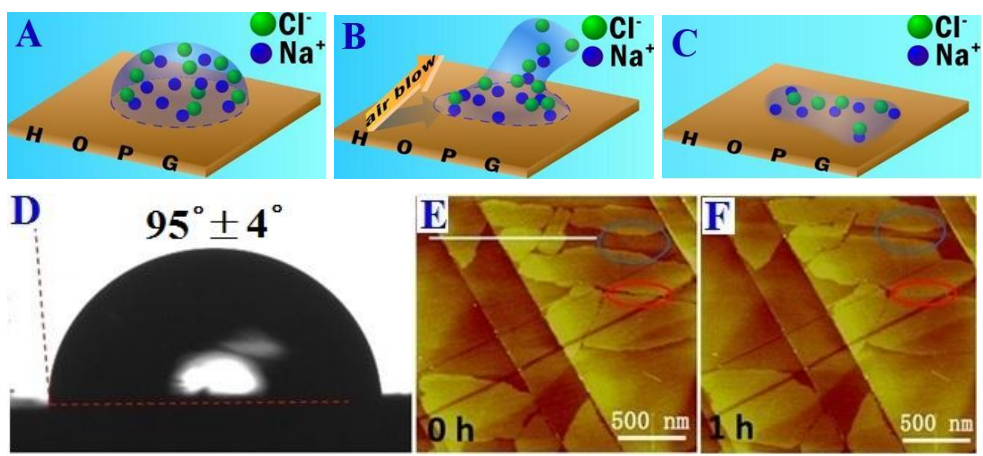

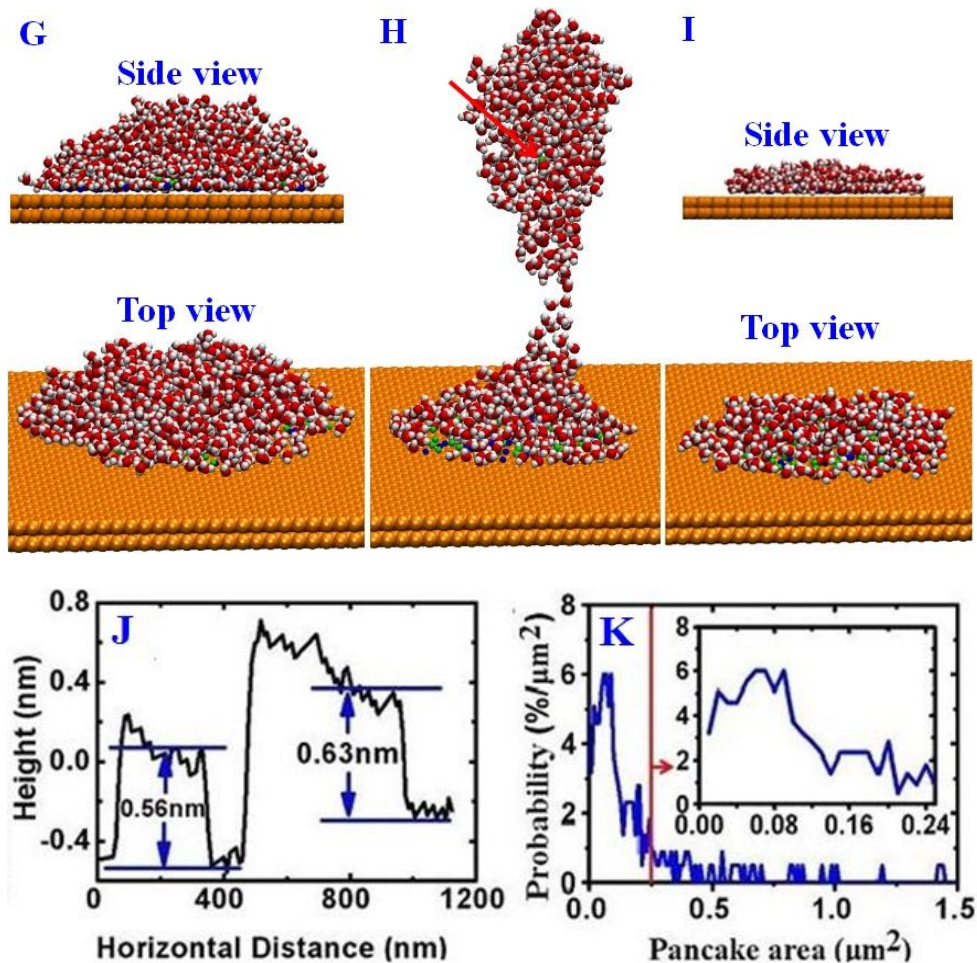

**Figure 1** | (A–C) Schematic drawings of the sample preparation in the experiment. (A) Deposition of a drop of NaCl solution onto a freshly cleaved highly oriented pyrolytic graphite (HOPG) sheet. (B) Solution partially removed by blowing air. (C) Resulting aqueous pancake (not drawn to scale). (D) A macroscopic droplet of aqueous salt solution on the HOPG surface. Average static contact angle (SCA) of NaCl solution on the HOPG surface is 95° ± 4°, which is close to the average SCA of 93° ± 2° of pure water. (E, F) AFM images of the HOPG surface at relative humidity of about 40%. (E) AFM image of the graphite surface after the aqueous droplet is removed by blowing air. The red ring indicates the boundary of a single pancake and follows its changes in shape over time. The blue ring indicates the formation of a bulge at the edge of the pancake. (F) One hour after treatment with salt solution, as shown in (E). The pancake has gradually enlarged and a visible bulge appears at the edge in the blue ring. (G–I) Snapshots from molecular dynamics simulations. (G) A NaCl solution droplet on a graphite surface. (H) A snapshot of the system when the salt solution is driven to the upper right; i.e., an acceleration of 0.10 nm/ps² was applied to all the water and ions along the midline between the x and z directions to study the effect of blowing air on the system during experimental preparation. (I) NaCl solution pancake on a graphite surface after water molecules were driven to the upper right for 2 ns. In

G, H and I, the orange structures depict the graphite sheets; water molecules, sodium and chlorine ions are shown with oxygen in red, hydrogen in white, sodium in blue and chlorine in green. (J) A height profile corresponding to the white line in (E) showing the layer is about 0.6 nm high relative to the substrate. (K) The distribution probability of micro pancake areas. The small figure in the upper-right corner is an enlargement of the left region marked by the red line.

Usually, many salt particles and aggregates were observed remaining on the graphite if the salt concentration was too high, and the surface was observed to be clean if the concentration was too low. However, at a salt concentration of ~20 mM, we observed many thin films of the salt solution (which we termed pancakes) on the HOPG surface approximately 10 min after treatment with the salt solution in most cases by tapping-mode atomic force microscopy (AFM) imaging as shown in Fig. 1(E) (see details in Supporting Information). The pancakes had apparent height of about 0.6 nm (Fig. 1(J)) in most cases and about 0.3 nm in some cases. The lateral scale of the pancakes spanned from a few hundred nanometers to several micrometers, which is about 3–4 orders of magnitude greater than the height. In one experiment, we determined the probability distribution of the areas of 217 pancakes that we observed (Fig. 1 (K)). From this distribution, we computed the average area as 0.20 $\mu m^2$. We found that pancakes formed at relative humidity ranging from 30%–70% and that no pancakes formed on surfaces treated only with pure water (see Supporting Information).

The pancakes appeared to move across the graphite surface over time. The AFM images of the surfaces acquired after 1 hour differ from those acquired 10 min after treatment with the salt solution (Fig. 1(F) vs. Fig. 1E). In Fig. 1(F), for example, the pancake has gradually enlarged and there is a visible bulge at the edge. The distance between pancakes decreased and the pancakes gradually coalesced with one another. These motion behaviors of pancakes enlarging and coalescing reduce the charge distribution and surface energy (see details in Supporting Information).

We also employed **non-contact**-mode vibrating scanning polarization force microscopy in imaging the HOPG surface treated by the salt solution to minimize the effect of AFM tips. As shown in Fig. S3 in the Supporting Information, the pancakes are again clearly seen to move. The motion of the pancakes was further demonstrated by the repair of a partially damaged pancake and the behavior of a thoroughly removed pancake (see Supporting Information). These observations suggest that the pancakes are liquid.

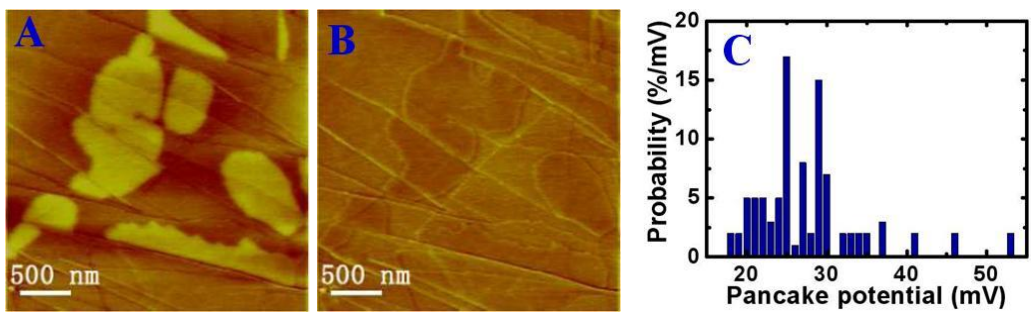

**Figure 2 | Electrical properties of the NaCl solution pancake.** Typical electrostatic force microscopy images obtained with tip voltage of 3 V (A) and −3 V (B). (C) The distribution probability of the electric potentials of 60 pancakes.

Remarkably, the salt-solution pancakes showed behavior that they were not charge neutral. The electrostatic force microscopy image obtained with a tip voltage of +3 V (Fig. 2(A)) showed clear and bright pancakes, and these pancakes darkened when the voltage was −3 V (Fig. 2(B)). From Kelvin probe force microscopy images, we obtained the potentials on the pancakes relative to the substrate. We measured the probability distribution of the electric potentials of 60 pancakes and determined the average pancake potential to be ~27 mV (Fig. 2, C).

The observed liquid pancakes on HOPG were unexpected since the HOPG surface is hydrophobic and the salt solution would be expected to form droplets (Fig. 1D). As the NaCl droplets dry, the left salt should form particles or aggregates of small particles on HOPG at 40% relative humidity, which is far below the deliquescence humidity of NaCl (around 75% relative humidity at room temperature[52]). This clearly indicates that the hydrophobic graphite surfaces had "apparent" and strong molecular-scale hydrophilicity with respect to the salt solution under ambient conditions.

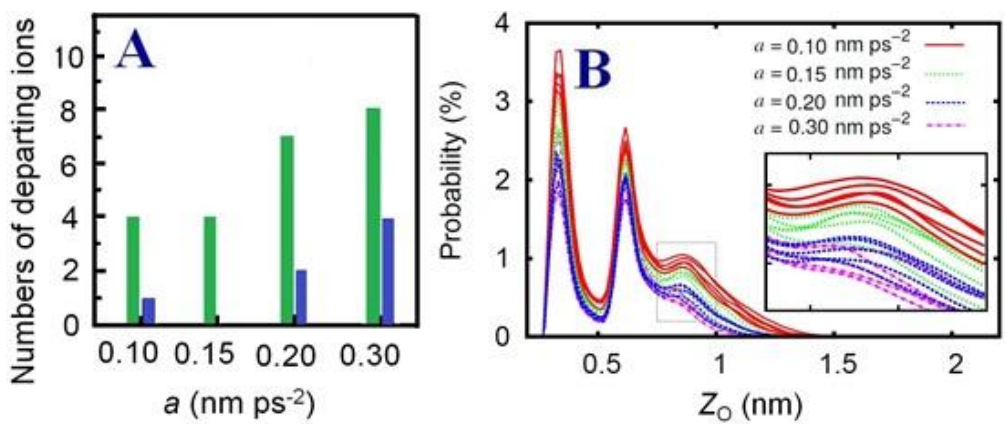

**Figure 3 |** (A) Total numbers of Cl⁻ ions (green ribbons) and Na⁺ ions (blue ribbons) in the solution separating from the graphite surface for five parallel samples at each acceleration value *a*.
corrections6

(B) Distribution probability of the oxygen atoms along the z direction in the resulting aqueous pancakes.

The key mechanism of the physics of these unexpected charged solution pancakes is the cation–π interactions. The graphite surface adsorbed ions, and water molecules were then adsorbed onto the surface with the help of the ions.[47] To further demonstrate the hydrophobic/hydrophilic transition due to the Na$^+$ adsorption onto graphite, we performed molecular dynamics (MD) simulations with a modification of the cation–π interaction[48] of the NaCl solution on graphite. We mimic these behaviors by applying additional acceleration *a* along the line between the x and z directions on a NaCl drop solution (comprising 810 water molecules and 45 sodium ions and 45 chloride ions) on the graphite surface. For each system, MD simulations were performed for five parallel samples with different initial configurations, with each simulation running for 2 ns. In a typical case shown in Fig. 1(G–I) where $a = 0.1$ nm/ps$^2$, at $t = 0.96$ ns, some of the solution moved upward and separated from the remaining solution (Fig. 1H) (see the movie in Supporting Information). Interestingly, there was a Cl$^-$ ion in the part of the solution departing the surface. The solution pancake remaining on the solid surface (see Fig. 1I) was thus positively charged because of the greater number of Na$^+$ ions. A variation of this charge behavior was observed by Martinez-Martin et al.[41] in the adsorption of atmospheric contaminants, such as a polycyclic aromatic hydrocarbon and its isomers, on graphitic surfaces.

Figure 3A shows the total numbers of Cl$^-$ and Na$^+$ ions in the solution separating from the graphite surface at different accelerations. There are fewer Na$^+$ ions than Cl$^-$ ions in the separated solution and, generally speaking, there are more Na$^+$ and Cl$^-$ ions in the separated solution with increasing acceleration. This positive charge behavior of the pancake on the graphite surface clearly results from the interactions between Cl$^-$ ions and the graphite surface being much weaker than the interactions between Na$^+$ ions and the graphite surface (i.e., the hydrated Cl$^-$–π interaction (–1.8 kcal/mol) is only about 1/10 of the hydrated Na$^+$–π interaction (–16.4 kcal/mol)[48]).

To study the behavior of the solution remaining on the graphite surface, we further performed MD simulations on the remaining NaCl solution for 4 ns after blowing away the separated solution. From the distribution probability of oxygen atoms along the z direction shown in Fig. 3B for each simulation, we see that there are three peaks at $Z_O$ = ~0.32, ~0.63 and ~0.86 nm. The first two peaks are consistent with the experimental observations of the height of the pancake structures of ~0.3 and ~0.6 nm. As *a* increases, the peak at $Z_O$ = ~0.86 nm reduces and even disappears. Only three and two simulations show clear peaks at $a$ = 0.2 and 0.3 nm/ps$^2$, respectively. Thus, the heights of the resulting pancakes depend on the acceleration *a*. We note that the existence of charged pancakes is robust as can be clearly seen at all four accelerations ($a$ = 0.10, 0.15, 0.20 and 0.30 nm/ps$^2$) in our simulations. This further demonstrates the consistency between the MD simulations and experimental observations.

We also investigated the main reason for the inconsistency in the wetting property between molecular and macroscopic scales. The flat graphite surface on the macroscopic scale is made up of a large number of graphene flakes stacking on a molecular scale. The macroscopic flat graphite surface is actually formed by a large number of molecular-level graphene layers, and there are many steps between the layers (see Fig. 1(E, F) and Fig. 2(A, B)). Theoretical calculations based on density functional theory (see details in Supporting Information) show that the Na$^+$ ions diffuse easily on the graphene flakes (barrier energies of only ~3 kcal/mol, Fig. S5B) but with more difficulty across the steps between layers (barrier energies exceeding 100 kcal/mol, Fig. S5B). Thus, Na$^+$ ions bind at the aromatic rings on the graphite surface retaining their hydration water molecules, resulting in the formation of molecular-thick pancakes of aqueous salt solution on the molecular-level graphene layers (see Fig. 1(E, F)), yet the drop of salt solution remains intact because there are many molecular-scale layer steps hindering the diffusion of Na$^+$ ions on the macroscopic surface.

Moreover, the many hydrophobic steps reduce the surface energy of the graphite surface (Fig. S6), resulting in the macroscopic graphite surface being more hydrophobic than the molecular-scale graphene flakes (see details in Supporting Information). It is much easier for the distribution of Na$^+$ ions in the salt solution on a molecular scale[48] to close the solid surface than that on a macroscopic scale,[53] which also affects the wetting property at different (macroscopic and molecular) scales. In summary, the different properties of the graphitic surface and Na$^+$ ion distributions between molecular and macroscopic scales produce the inconsistency in the wetting property between molecular (hydrophilicity) and macroscopic (hydrophobicity) scales.

**CONCLUSION**

We conclude that, counter to intuition, molecular-thick films of aqueous salt solution can stably exist on a hydrophobic carbon-based surface under ambient conditions. Experimental and theoretical results show the unexpected molecular-scale hydrophilicity on hydrophobic surfaces. The cation–π interactions and the different properties of the graphitic surface and Na$^+$ ion distributions between molecular and macroscopic scales generate an inconsistency in the wetting property between molecular (hydrophilicity) and macroscopic (hydrophobicity) scales. Interestingly, the formed pancakes spontaneously display positively charged behavior. Considering that the key ingredient in the production of molecular-scale pancakes, the aromatic rings, is commonly found in biomolecules, the findings may clarify the actual interactions of biomolecules. The underlying mechanism should be helpful in understanding and controlling the functional characteristics of carbon-based materials for various applications such as drug delivery, water purification employing carbon nanotubes, ion filtration employing graphene pores, hydrogen storage employing graphene/graphite, and other applications of carbon-based nanomaterials.

## MATERIALS AND METHODS

**Experimental Section:**

**Materials.** NaCl (crystal purity 99.99%) was purchased from Sinopharm Chemical Reagent Co. Ltd., and was dissolved in ELGA laboratory water to a final concentration of 20 mM. The highly oriented pyrolytic graphite (HOPG) was provided by Molecular Devices and Tools for Nano Technology Co. Zelenograd, Moscow, Russia. The electric conductive adhesive (DAD-40) was purchased from the Shanghai Research Institute of Synthetic Resins.

**Sample Preparation.** The HOPG fragment was fixed to the sample holder using electric conductive adhesive with the working side up, and was freshly cleaved using double-faced adhesive tape. The NaCl sample was prepared according to a process previously reported for the observation of liquid nanodroplets of KOH by scanning polarization force microscopy[54] on graphite. Briefly, a drop (~20 μl) of NaCl solution (20 mM) was deposited on the HOPG substrate, and the droplet was then dried with a stream of air at room temperature and 40%–60% relative humidity. The relative humidity and temperature of a sample chamber (SDH-01N, Shanghai Jianheng Instrument Co.) were controlled with accuracy of 5% and 0.1 ℃, respectively. The as-prepared sample was placed in this chamber for a given amount of time.

**Atomic Force Microscope (AFM) Imaging.** Experiments were performed on a commercial AFM (Nanoscope IIIa, Veeco/Digital Instruments, Santa Barbara, CA, USA) equipped with a J scanner (100 μm × 100 μm) and E scanner (15 μm × 15 μm). Silicon etched probes (NSC18/Ti-Pt, MikroMasch Co., length: 230 μm, width: 40 μm, thickness: 3 μm, nominated spring constant: 3.5 N/m, resonant frequency: 60–90 kHz) were used in other experiments. The Ti-Pt coating comprised a 10-nm Pt layer on a 20-nm Ti sublayer, which provided greater adhesion and electromigration firmness than if using Pt alone. The Ti-Pt coating formed on both the tip and reflective side of the cantilever. The resulting tip radius with the coating was 40 nm. The morphological features of the AFM images, namely the height and width, were analyzed using AFM accessory software (ver. 7.30). All AFM images were adequately flattened using the software to correct the distortion at a micrometer scale, but no other digital operation was carried out. All AFM data were obtained at room temperature, whereas relative humidity was measured by a hygrometer with accuracy of 5% (SDH-01N, Shanghai Jianheng Instrument Co.).

**SCA Measurement.** The SCA measurement was made on an Attension Theta system (KSV Instruments Ltd., Finland). The volume of each droplet of NaCl solution or pure water was ~5 μl and each droplet was carefully touched to the sample surface. A digital camera was used to take images of all droplets, and the values of SCA were automatically computed by the supplied calculation software. Each HOPG sample was measured at three different points and the average value was reported.

**Computational Methods.** The cation–π interactions between Na+ and the graphite surface are represented by the model potential[48]

$$V = \varepsilon ((z_m/z)^8 - 2 (z_m/z)^4), \qquad (1)$$

where the parameters $\varepsilon$ and $z_m$ are the adsorption energy and balance position (the vertical dimension between the $Na^+$ ion and the surface) of $Na^+$ relative to the graphite surface and $z$ is the vertical distance between the $Na^+$ ion and the surface; these are the main potential parameters describing the cation–π interaction. Their values are $z_m$ = 3.8 Å and $\varepsilon = \varepsilon_0 = -16.4$ kcal/mol.[48] Molecular dynamics simulations were carried out using the program NAMD2/VMD1.9 packages,[55] with the CHARMM force field,[56] at time steps of 2 fs and with the O–H bonds and C atoms held fixed (see detail in Supporting Information).

REFERENCES

[1] Hu, J., Xiao, X. D., Ogletree, D. F. & Salmeron, M. Imaging the condensation and evaporation of molecularly thin films of water with nanometer resolution. *Science* **268**, 267-269 (1995).

[2] Xu, K., Cao, P. & Heath, J. R. Graphene visualizes the first water adlayers on mica at ambient conditions. *Science* **329**, 1188-1191 (2010).

[3] Bonn, D., Eggers, J., Indekeu, J., Meunier, J. & Rolley, E. Wetting and spreading. *Rev. Mod. Phys.* **81**, 739-805 (2009).

[4] Feibelman, P. J. The first wetting layer on a solid. *Phys. Today* **63**, 34-39 (2010).

[5] Cicero, G., Calzolari, A., Corni, S. & Catellani, A. Anomalous wetting layer at the Au(111) surface. *J. Phys. Chem. Lett.* **2**, 2582-2586 (2011).

[6] Shakhnovich, E. Protein folding thermodynamics and dynamics: where physics, chemistry, and biology meet. *Chem. Rev.* **106**, 1559-1588 (2006).

[7] Berne, B. J., Weeks, J. D. & Zhou, R. H. Dewetting and hydrophobic interaction in physical and biological systems. *Annu. Rev. Phys. Chem.* **60**, 85-103 (2009).

[8] Wu, Z., Cui, Q. & Yethiraj, A. Driving force for the association of hydrophobic peptides: The importance of electrostatic interactions in coarse-grained water models. *J. Phys. Chem. Lett.* **2**, 1794-1798 (2011).

[9] Bier, D. et al. Molecular tweezers modulate 14-3-3 protein–protein interactions. *Nat. Chem.* **5**, 234-239 (2013).

[10] Král, P. & Wang, B. Material drag phenomena in nanotubes. *Chem. Rev.* **113**, 3372-3390 (2013).

[11] Mulvery, J. J. et al. Self-assembly of carbon nanotubes and antibodies on tumours for targeted amplified delivery. *Nat. Nanotech.* **8**, 763-771 (2013).

[12] Majumder, M., Chopra, N., Andrews, R. & Hinds, B. J. Nanoscale hydrodynamics: Enhanced flow in carbon nanotubes. *Nature* 438, 44-44 (2005).

[13] Powell, M R., Cleary, L., Davenport, M., Shea, K. J. & Siwy, Z. S. Electric-field-induced wetting and dewetting in single hydrophobic nanopores. *Nat. Nanotech.* **6**, 798-802 (2011).

[14] Contreras, F.-X. et al. Molecular recognition of a single sphingolipid species by a protein's transmembrane domain. *Nature* **481**, 525-529 (2012).

[15] Chandler, D. Interfaces and the driving force of hydrophobic assembly. *Nature* **437**, 640 (2005).

[16] Law, A. D., Auriol, M., Smith, D., Horozov, T. S. & Buzza, D. M. A. Self-assembly of two-dimensional colloidal clusters by tuning the hydrophobicity, composition, and packing geometry. *Phys. Rev. Lett.* **110**, 138301 (2013).

[17] Baram, M., Chatain, D. & Kaplan, W. D. Nanometer-thick equilibrium films: the interface between thermodynamics and atomistics. *Science* **332**, 206 (2011).

[18] Coridan, R. H. et al. *Phys. Rev. Lett.* **103**, 237402 (2006).

[19] Wang, S. et al. Enthalpy-driven three-state switching of a superhydrophilic/superhydrophobic surface. *Angew. Chem. Int. Edit.* **46**, 3915-3917 (2007).

[20] Zhu, C., Li, H., Huang, Y., Zeng, X. C. & Meng S. Microscopic insight into surface wetting: Relations between interfacial water structure and the underlying lattice constant. *Phys. Rev. Lett.* **110**, 126101 (2013).

[21] Wang, C. et al. Stable liquid water droplet on water monolayer formed at room temperature on ionic model substrates. *Phys. Rev. Lett.* **103**, 137801 (2009).

[22] Bai, J. & Zeng, X. C. Polymorphism and polyamorphism in bilayer water confined to slit nanopore under high pressure. *Proc. Natl. Acad. Sci. USA* **109**, 21240-21245 (2012).

[23] Liu, J., Wang, C., Guo, P., Shi, G. & Fang, H. Linear relationship between water wetting behavior and microscopic interactions of super-hydrophilic surfaces. *J. Chem. Phys.* **139**, 234703 (2013).

[24] Tu, Y. et al. Destructive extraction of phospholipids from Escherichia colimembranes by graphene nanosheets. *Nat. Nanotechnol.* **8**, 594-601 (2013).

[25] Li, H. & Zeng X. C. Two dimensional epitaxial water adlayer on mica with graphene coating: An *ab initio* molecular dynamics study. *J. Chem. Theory. Comput.* **8**, 3034-3043 (2012).

[26] Gao, Y., Shao, N., Zhou, R., Zhang, G. & Zeng, X. C. [CTi7 2+]: Heptacoordinate carbon motif? *J. Phys. Chem. Lett.* **3**, 2264-2268 (2012).

[27] Chen, D., Feng, H., & Li, J. Graphene oxide: Preparation, functionalization, and electrochemical aplications. *Chem. Rev.* **112**, 6027-6053 (2012).

[28] Grorgakilas, V. et al. Functionalization of graphene: Covalent and non-covalent approaches, derivatives and applications. *Chem. Rev.* **112**, 6156-6214 (2012).

[29] Hu, L., Hecht, D. S. & Grüner, G. Carbon nanotube thin films: Fabrication, properties, and applications. *Chem. Rev.* **110**, 5790-5844 (2010).

[30] Dillon, A. C. Carbon nanotubes for photoconversion and electrical energy storage. *Chem. Rev.* **110**, 6856-6872 (2010).

[31] Thilgen, C. & Diederich, F. Structural aspects of fullerene chemistry－A journey through fullerene chirality. *Chem. Rev.* **106**, 5049-5135 (2006).

[32] Manyà, J. J. Pyrolysis for biochar purposes: A review to establish current knowledge gaps and research needs. *Environ. Sci. Technol.* **46**, 7939-7954 (2012).

[33] Novak, J. M. et al. Characterization of designer biochar produced at different temperatures and their effects on a loamy sand. *Ann. Environ., Sci.* **3**, 195-206 (2009).

[34] Mehta, B. A., Nelson, E. J., Webb, S. M. & Holt, J. K. The interaction of bromide ions with graphitic materials. *Adv. Mater.* **21**, 102-106 (2009).

[35] Mahadevi, A. S. & Sastry, G. N. Cation−π interaction: Its role and relevance in chemistry, biology, and material science. *Chem. Rev.* **113**, 2100−2138 (2013).

[36] Kogut, B. M. Assessment of the Humus Content in Arable Soils of Russi. *Eur. Soil Sci.* **35**, 843-851 (2012).

[37] Hitzel, A., Pöhlmann, M., Schwägele, F., Speer, K. & Jira, W. Polycyclic Aromatic Hydrocarbons (PAH) and Phenolic Substances in Meat Products Smoked with Different Types of Wood and Smoking Spices. *Food Chem.* **139**, 955-962 (2013).

[38] Rafiee, J. F. et al. Wetting transparency of graphene. *Nat. Mater.* **11**, 217-222 (2012).

[39] Shih, C. et al. Breakdown in the wetting transparency of graphene. *Phys. Rev. Lett.* **109**, 176101 (2012).

[40] Li, Z. et al. Effect of airborne contaminants on the wettability of supported graphene and graphite. *Nat. Mater.* **12**, 925-931 (2013).

[41] Martinez-Martin, D. et al. Atmospheric contaminants on graphitic surfaces. *Carbon* **61**, 33-39 (2013).

[42] Shi, G., Yang, J., Ding, Y. & Fang, H. Orbital effect-induced anomalous anion–π interactions between electron-rich aromatic hydrocarbons and fluoride. *ChemPhysChem* **15**, 2588-2594 (2014).

[43] Shi, G., Ding, Y. & Fang, H. Unexpectedly strong anion–π interactions on the graphene flakes. *J. Comput. Chem.* **33**, 1328-1337 (2012).

[44] Yang, J., Shi, G., Tu, Y. & Fang, H. High correlation between oxidation loci on graphene oxide. *Angew. Chem. Int. Edit.* **53**, 10190-10194 (2014).

[45] Patra, N., Esan, D. A. & Král, P. Dynamics of ion binding to graphene nanostructures. *J. Phys. Chem. C* **117**, 10750-10754 (2013).

[46] Sunner, J., Nishizawa, K. & Kebarle, P. Ion-solvent molecule interactions in the gas phase. The potassium ion and benzene. *J. Phys. Chem.* **85**, 1814-1820 (1981).

[47] Shi, G., Wang, Z., Zhao, J., Hu, J. & Fang, H. Adsorption of sodium ions and hydrated sodium ions on the hydrophobic graphite surface via cation-π interactions. *Chin. Phys. B* **20**, 068101 (2011).

[48] Shi, G. et al. Ion enrichment on the hydrophobic carbon-based surface in aqueous salt solutions due to cation-π interactions. *Sci. Rep.* **3**, 3436 (2013).

[49] Dougherty, D. A. The cation-π interaction. *Acc. Chem. Res.* **46**, 885-893 (2013).

[50] Daze, K. D. & Hof, F. The cation-π interaction at protein-protein interaction interfaces: developing and learning from synthetic mimics of proteins that bind methylated lysines. *Acc. Chem. Res.* **46**, 937-945 (2013).

[51] Duan, M. et al. Cation⊗3π: cooperative interaction of a cation and three benzenes with an anomalous order in binding energy. *J. Am. Chem. Soc.* **134**, 12104-12109 (2012).

[52] Hucher, M., Oberlin, A. & Hocart, R. Adsorption de vapeur d'eau sur les faces de clivage de quelques halogénures alcalins. *Bull. Soc. Fr. Mineral. Cristallogr.* **90**, 320-332 (1967).

[53] Garrett, B. C. Ions at the Air/Water Interface. Science 303, 1146–1147 (2004).

[54] Hu, J., Carpick, R. W., Salmeron, M. & Xiao, X. D. Imaging and manipulation of nanometer-size liquid droplets by scanning polarization force microscopy. *J. Vac. Sci. Technol. B* **14**, 1341-1343 (1996).

[55] Phillips, J. C. et al. Scalable Molecular Dynamics with NAMD. *J. Comput. Chem.* **26**, 1781-1802 (2005).

[56] MacKerell, A. D. et al. All-Atom Empirical Potential for Molecular Modeling and Dynamics Studies of Proteins. *J. Phys. Chem. B* **102**, 3586-3616 (1998).

**ACKNOWLEDGMENTS**

We thank Drs. Yi Zhang, Jiang Li, Jingye Li, Qing Ji and Jijun Zhao for their constructive suggestions. This work was supported by the National Science Foundation of China (under grant nos. 11290164, 11404361 and 11204341), the Shanghai Natural Science Foundation of China (under grant no. 13ZR1447900), the Knowledge Innovation Program of SINAP, the Deepcomp7000 and ScGrid of Supercomputing Center, Computer Network Information Center of Chinese Academy of Sciences and the Shanghai Supercomputer Center of China.

**Author contributions**

G.S. and J.L. performed molecular dynamics simulations. H.F. and G.S. carried out most of the theoretical analysis. Y.S. and J.H. designed and observed the experimental investigation. C.W. and B.S. carried out some theoretical analysis. Y.W. carried out some experimental analysis. H.F., J.H. and G.S. contributed most of the ideas and wrote the paper. All authors discussed the results and commented on the manuscript.

**Supporting Information**

Supplementary information accompanies this paper at http://www.nature.com/scientificreports.

Competing financial interests: The authors declare that they have no competing financial interests.